\documentclass[reprint,superscriptaddress,amsmath,amssymb,aps,pra,floatfix]{revtex4-2}
\pdfoutput=1
\usepackage[utf8]{inputenc}
\usepackage[british]{babel} 

\usepackage{textgreek} 

\usepackage{dcolumn}
\usepackage{bm}
\usepackage{physics}

\usepackage{graphicx}
\graphicspath{ {images/} }
\usepackage[caption=false]{subfig}

\usepackage{enumerate}
\usepackage[shortlabels]{enumitem}

\usepackage{nccmath} 
\usepackage{SIunits}
\usepackage{hyphenat}
\usepackage{hyperref}
\usepackage[capitalise]{cleveref}
\crefname{section}{Sec.}{Secs.}
\Crefname{section}{Section}{Sections}
\usepackage{braket}
\usepackage{floatrow}

\newcommand{\psnspd}{P\nobreakdash-SNSPD}

\begin{document}


\title{High-efficiency and fast photon-number resolving parallel superconducting nanowire single-photon detector}

\author{Lorenzo~Stasi}
\email{Corresponding author: lorenzo.stasi@idquantique.com}
\affiliation{ID Quantique SA, CH-1227 Carouge, Switzerland}
\affiliation{Group of Applied Physics, University of Geneva, CH-1211 Geneva, Switzerland}

\author{Ga\"etan~Gras}
\affiliation{ID Quantique SA, CH-1227 Carouge, Switzerland}

\author{Riad~Berrazouane}
\affiliation{ID Quantique SA, CH-1227 Carouge, Switzerland}

\author{Matthieu~Perrenoud}
\affiliation{Group of Applied Physics, University of Geneva, CH-1211 Geneva, Switzerland}

\author{Hugo~Zbinden}
\affiliation{Group of Applied Physics, University of Geneva, CH-1211 Geneva, Switzerland}

\author{F\'elix~Bussi\`eres}
\affiliation{ID Quantique SA, CH-1227 Carouge, Switzerland}

\date{\today}

\begin{abstract}
Photon-number resolving (PNR) single-photon detectors are an enabling technology in many areas such as photonic quantum computing, non-classical light source characterisation and quantum imaging. Here, we demonstrate high-efficiency PNR detectors using a parallel superconducting nanowire single-photon detector (P-SNSPD) architecture that does not suffer from crosstalk between the pixels and that is free of latching. The behavior of the detector is modelled and used to predict the possible outcomes given a certain number of incoming photons. We apply our model to a 4-pixel P-SNSPD with a system detection efficiency of 92.5\%.  We also demonstrate how this detector allows reconstructing the photon-number statistics of a coherent source of light, which paves the way towards the characterisation of the photon statistics of other types of light source using a single detector.
\end{abstract}

\maketitle

\section{Introduction}
Quantum states of light are made of superposition of photon number states. This is at the core of quantum optics and all of its practical applications such as quantum key distribution with coherent states~\cite{gisin2002quantum, pirandola2020advances,xu2020secure},linear optical quantum computing based on downconverted photon pairs or on squeezed states of light~\cite{slussarenko2019photonic}.
In this context, photon-number resolving (PNR) detectors are known to play a key role in the processing itself~\cite{knill2001scheme}.
PNR detectors are also beneficial in several other fields such as the characterization of quantum sources of light~\cite{von2019quantum} and imaging with threshold detectors~\cite{cohen2019thresholded}. PNR detectors have been realized using different platforms~\cite{hadfield2016superconducting}.
Among these, transition-edge sensors (TES)~\cite{rosenberg2005noise} have so far demonstrated impressive performances in terms of combined efficiency and single-shot measurement fidelity~\cite{lita2008counting, morais2020precisely} thanks to their bolometric working principle. Photon-number resolution with TES however has shortcomings such as long recovery times ($\sim\micro\second$), high jitters ($\sim\nano\second$) and very low operating temperatures (between 50 to 100~mK). This currently prevents their use in high-repetition rate experiments and impairs the scalability of optical quantum processors.

Interestingly, these shortcomings can potentially be overcomed by superconducting nanowire single-photon detectors (SNSPDs), as they can have short recovery times ($\leq~10$~ns)~\cite{perrenoud2021operation} and low jitters (as low as a few~ps)~\cite{esmaeil2020efficient,caloz2018high, korzh2020demonstration}.
They can also have near unity efficiency~\cite{reddy2020superconducting} and very low dark count rates. Due to their working principle based on hot spot creation, SNSPDs do not however possess intrinsic photon-number resolution capability in the same way TES do. To achieve some degree of resolution, one has to resort to other approaches. One is to exploit the fact that several simultaneous hot spots can change the detector signal's slew rate~\cite{cahall2017multi,endo2021quantum} or amplitude when using impedance-matched tapers~\cite{zhu2019resolving}. These methods can yield discrimination up to 3-photon events and can help to discriminate one versus more than one photon pulses to improve the autocorrelation function of heralded photons \cite{davis2021improved, sempere2022reducing}. However, their resolution power quickly fades because with three or more photons absorbed, the overlap of the different signals makes the photon-number discrimination impossible.

To get more information about the number of absorbed photons, one can exploit multiplexing, either spatially with multi-pixel SNSPDs or temporally with delay lines~\cite{dauler2007multi, fitch2003photon}. Spatial multiplexing with several pixels to approach linear photon-number resolution, each connected to its own coaxial lines was demonstrated recently~\cite{zhang2020approaching}. This can be simplified using a parallel SNSPD (\psnspd{}) design where the pixels are connected in parallel to a single coaxial line, and the signal's amplitude informs on the number of pixels that clicked~\cite{divochiy2008superconducting,marsili2009superconducting, moshkova2019high}.
This \psnspd{} approach provide a path towards high-efficiency, low-jitter and short recovery time PNR detectors, as it offers a mean to probe the statistics of quantum light using a single device. Ultimately, \psnspd{}s have the potential to allow single-shot high-fidelity identification of the incident photon number approaching~\cite{provaznik2020benchmarking}.
Here we take a step in this direction by showing a high-performance \psnspd{} device and by using it to probe and reconstruct the statistics of quantum light. 

\begin{figure}
    \includegraphics[width=0.6\linewidth]{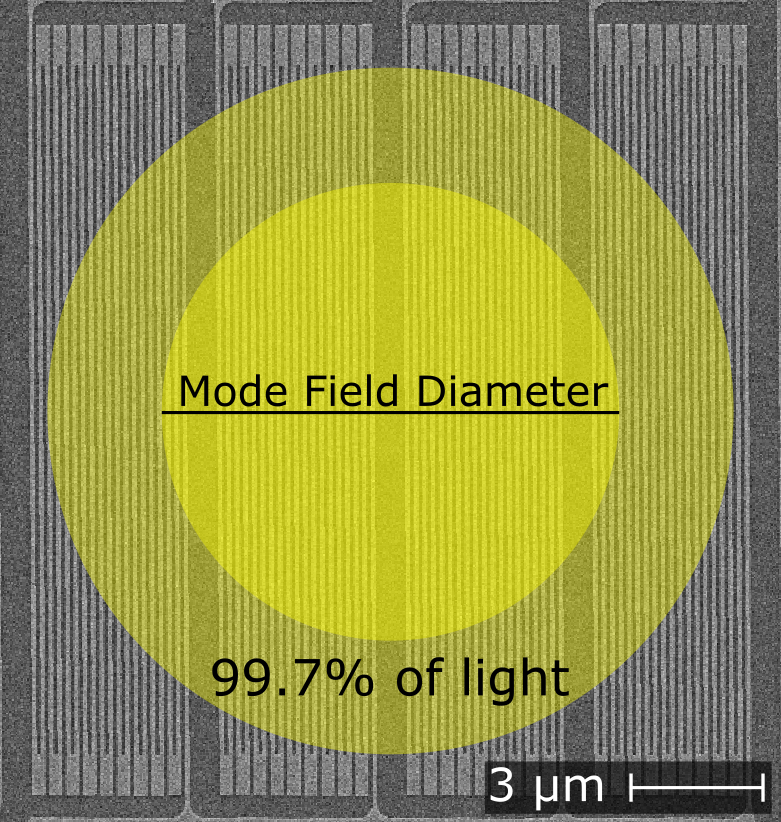}
    \caption{SEM images of a 4-pixels P-SNSPDs with the mode field diameter dimension of single-mode fiber on top.}
    \label{fig:SEM_P-SNSPDs}
\end{figure}

\section{P-SNSPD model}
A \psnspd{} is a set of individual SNSPD pixels that are connected in parallel. They are biased by the same current source and are illuminated such that each pixel receives a portion of the incident light, and each pixel has its own detection efficiency. When one or several pixels click simultaneously, the output signal's amplitude is proportional to the number of pixels that clicked. Here, we assume that the amplitudes are discrete and perfectly distinguishable, and that there is no instantaneous cross-talk between them.
We wish to develop a model of a \psnspd{} that can allow one to map the input statistics of the light to the output signals produced by the device. This has been attempted previously in Refs.~\cite{marsili2009superconducting, moshkova2019high} where assumptions such as the same efficiency for each pixel or a uniform spacial distribution of light were taken. These assumptions generally do not hold in practice, but are important to achieve the reconstruction of input light statistics. Another approach is to use quantum detector tomography~\cite{schapeler2020quantum,endo2021quantum} where the underlying details of the device does not need to be modelled. While this approach can work, it can also occult some interesting details about the device itself, such as different efficiencies between the pixels. The approach we take here does not make any prior assumption on the pixel efficiencies but allows us to estimate them. Furthermore, our approach requires only one set of data to fully characterize the device and is fast to implement.

Let us consider a \psnspd{} illuminated by a photon-number distribution that we write as a column vector $\mathbf{S}$, where each element is written as ${S{\left(m\right)}}$ and is the probability to have $m$-photons incident on the detector. Let $\mathbf{Q}$ be a column vector with elements $Q(n)$ representing the probability to observe each of the possible discrete amplitudes of the output signal, where $n = \{0, 1, \ldots, N\}$ with $N$ equal to the number of pixels. We denote the amplitude of $n$ pixels clicking as an $n$-click. We wish to find the matrix $\mathbf{P}$ that maps the input photon-number distribution \textbf{S} into the $n$-click probability distribution \textbf{Q} through the relation $\mathbf{Q} = \mathbf{P}\mathbf{S}$. Element wise, the relation is expressed as
\begin{equation}
    {Q(n)} = \sum_{m=0}^{\infty} P_{nm} \cdot S(m)
    \label{eq:povm}
\end{equation}
where ${P_{nm}}$ is the probability of registering an $n$-click when $m$ photons are incident. We note that the probabilities ${P_{nm}}$ can be used to define the elements of the positive-operator-valued measures (POVMs) of the detector~\cite{lundeen2009tomography}.  

While the $m$ can in principle take an infinite value, in practice it can be truncated to finite values. All the $P_{nm}$ elements with $n>m$ are set to~0, and if we truncate $m$ to the value $M$, then \textbf{P} takes the form:
\begin{equation}
    \mathbf{P} = \begin{bmatrix}
        P_{00} & P_{01} & P_{02} & \dots & P_{0M} \\
        0 & P_{11} & P_{12} & \dots & P_{1M} \\
        \vdots & \vdots & \ddots & \dots & \vdots \\
        0 & 0 & 0 & P_{nn} & P_{nM} \\
    \end{bmatrix}.
    \label{POVM_}
\end{equation}
To construct each element of the matrix $\mathbf{P}$, we enumerate all the possible cases as if the photons in the incident pulse are registered one at a time. Thus, for a specific $m$-photons with $n$-click event, we take into account two kinds of combinations: which photons are absorbed in the group of $m$, and which pixels detected them. 

As free parameters, we use the single pixel efficiencies $\eta_i$, which are composed by the product of the internal quantum efficiency and the coupling efficiency between the fiber and the pixel. The latter is needed in order to consider that each pixel could receive a different portion of light with respect to the others. See \cref{app:appA} for more details.

\section{Results}
We fabricated a 4-pixel \psnspd{} based on amorphous MoSi as the superconducting material using a process outlined in Ref.~\cite{perrenoud2021operation}. The particular architecture presented in this paper prevents the electrical crosstalk. The thermal crosstalk is avoided by engineering a gap between the pixels.

\begin{figure}[h]
    \centering 
    \includegraphics[width=0.9\linewidth]{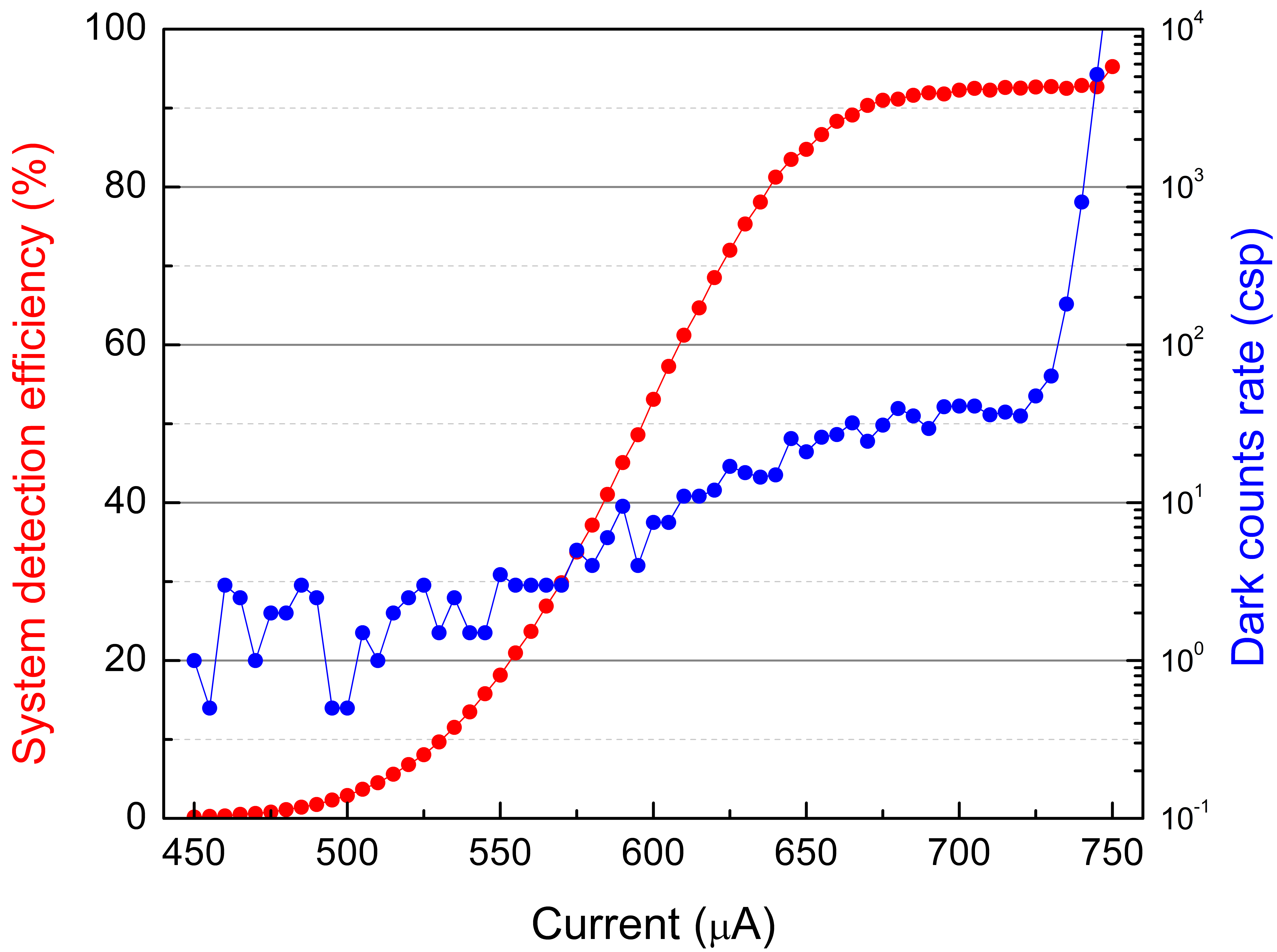}
    \caption{System detection efficiency and system dark counts for a 4-pixel PNR detector.}
    \label{fig:eff}
\end{figure}

We characterize the system detection efficiency (SDE) in a 3-stage cryostat at 0.8~K by using a calibrated powermeter and three variable digital attenuators. The light polarization is oriented in order to maximize the SDE. In~\cref{fig:eff}, the SDE \textit{vs} bias current is shown at a detection rate of $\sim$100 kHz. The presence of a plateau indicates the saturated internal quantum efficiency of the device and the maximum value is $92.5\pm2.4\%$. Dark counts rate (DCR) remains negligible along the plateau, around 140~cps. The DCR of the system alone, namely without the fiber plugged into the cryostat, are around 35~cps.

\begin{figure}[h] 
    \centering
    \includegraphics[width=0.85\linewidth]{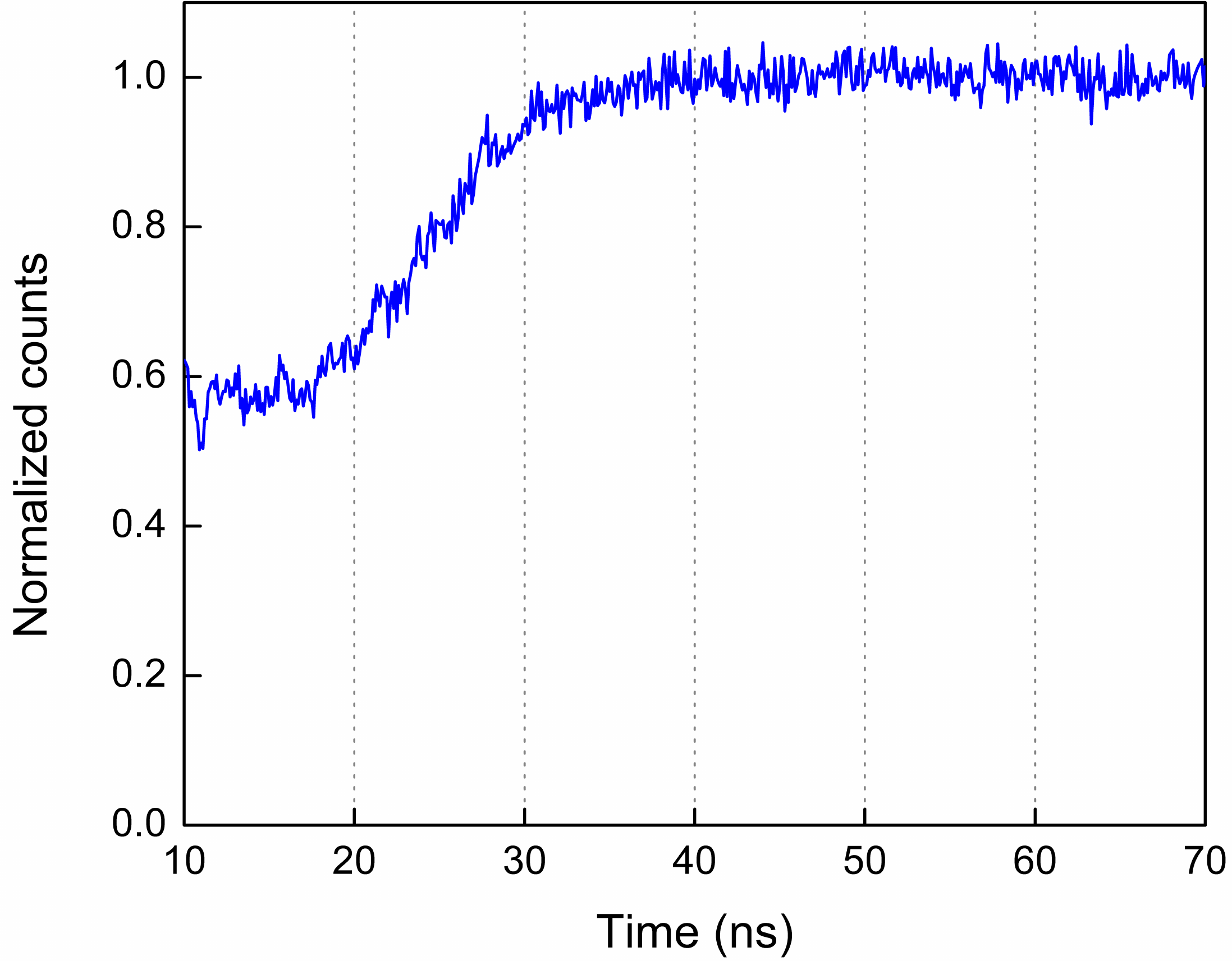}
    \caption{Recovery time of the 4 pixel PNR detector.}
    \label{fig:recovery}
\end{figure}

The detector recovery time (RT) (see \cref{fig:recovery}) is characterized by shining faint CW light on it and collecting all the registered clicks the detector generates. 
After a single-photon absorption by one of the pixels, the latter becomes inactive, but the overall device is still able to detect new incoming photons due to the three other active pixels. In fact, the \psnspd{} displays more than $50\%$ of its nominal efficiency after $10~\nano\second$ and it is back at full efficiency in $40~\nano\second$. 

\begin{figure}
    \centering
    \includegraphics[width=0.85\linewidth]{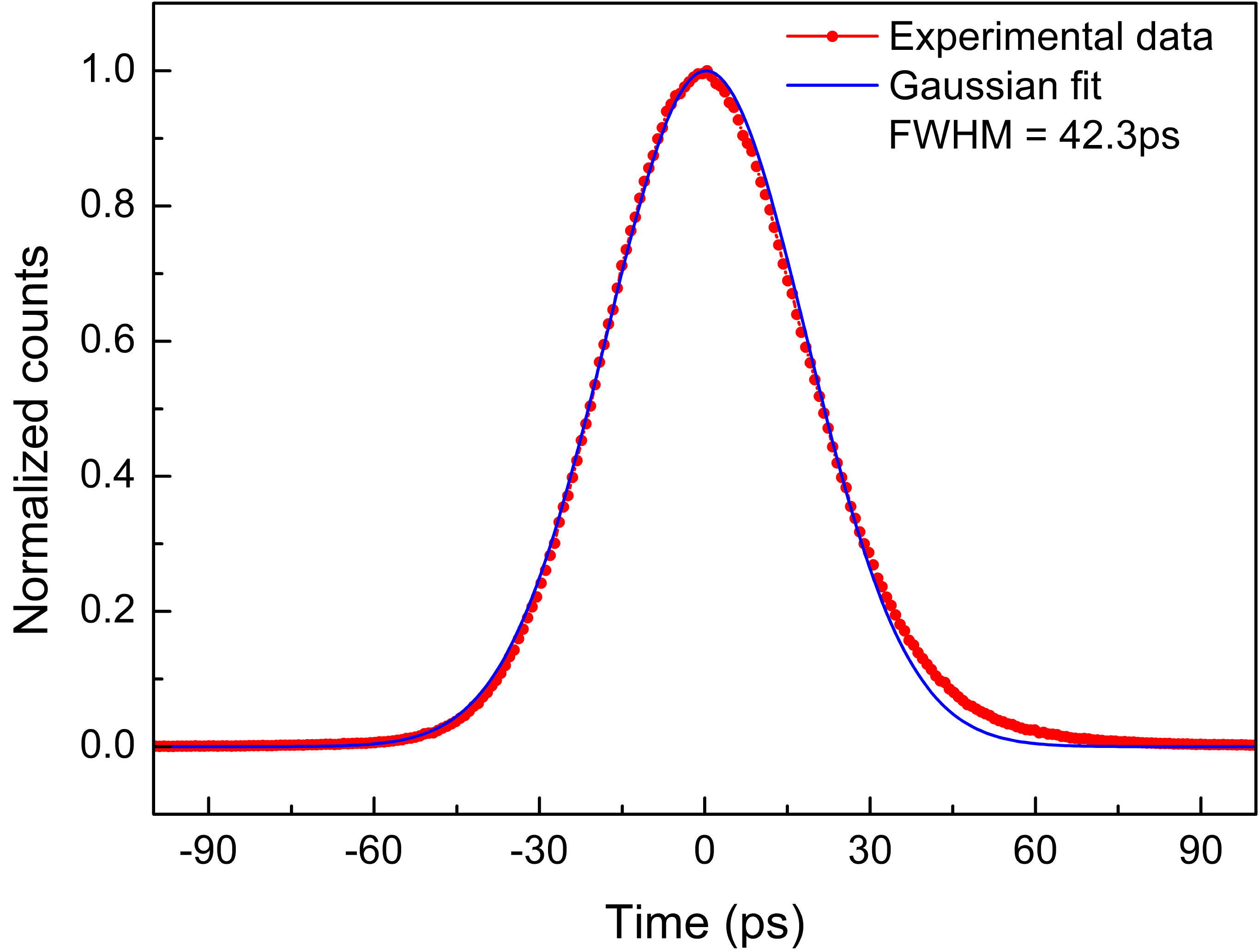}
    \caption{Jitter of the 4 pixel PNR detector.}
    \label{fig:jitter}
\end{figure}

Another important feature of SNSPDs is their timing jitter. 
For the P-SNSPDs, we obtain a jitter of 42~ps at full-width half-maximum (FWHM) at the single-photon level (see \cref{fig:jitter}), a relatively higher result with respect to state-of-the-art SNSPDs. The reason is due to the current redistribution among the outer pixels, which cause less current into the read-out circuit reducing the signal-to-noise ratio and thus increasing the jitter.

\begin{figure*}[t]
    \centering
    \includegraphics[width = 0.9\linewidth]{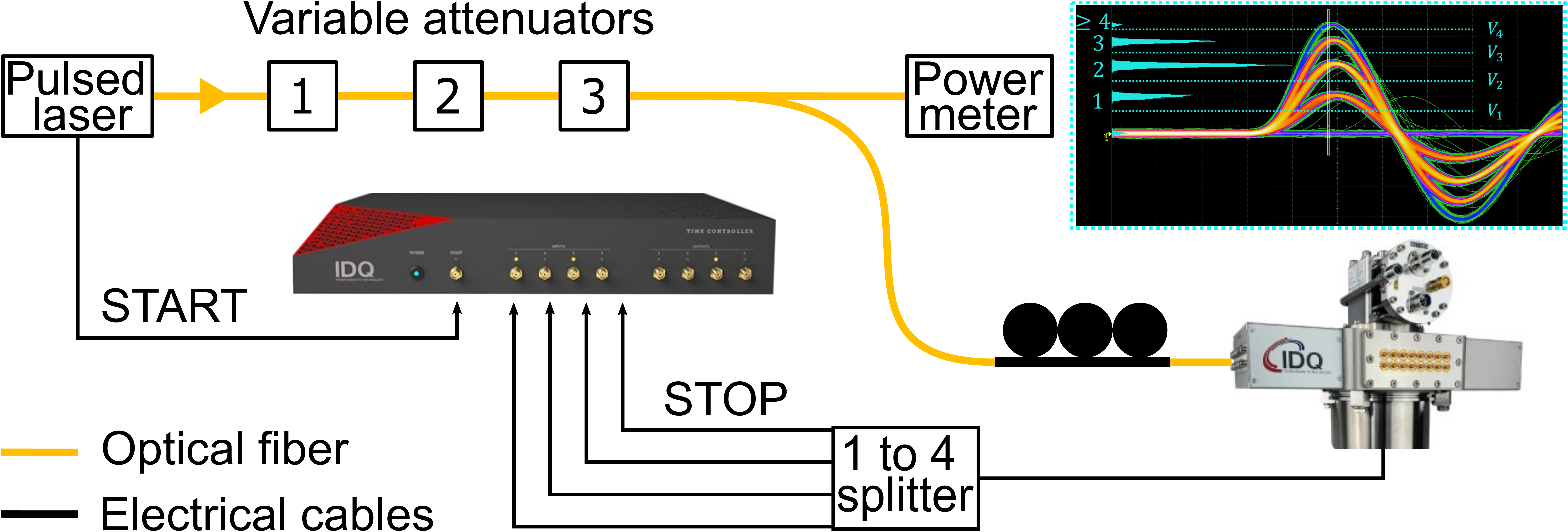}
    \caption{Schematic of the experiment's setup. Pulsed light is sent to the detector and the electrical signal produced is sent to a 1-to-4 resistive splitter before going to the time-tagger machine. On each input, a different threshold was set to obtain the corresponding photon-number event as it can be seen on the top-right inset.}
    \label{fig:setup}
\end{figure*}

\begin{figure*}[t]
    \centering
    \begin{minipage}{0.3\textwidth}
        \centering
        \includegraphics[width=\textwidth]{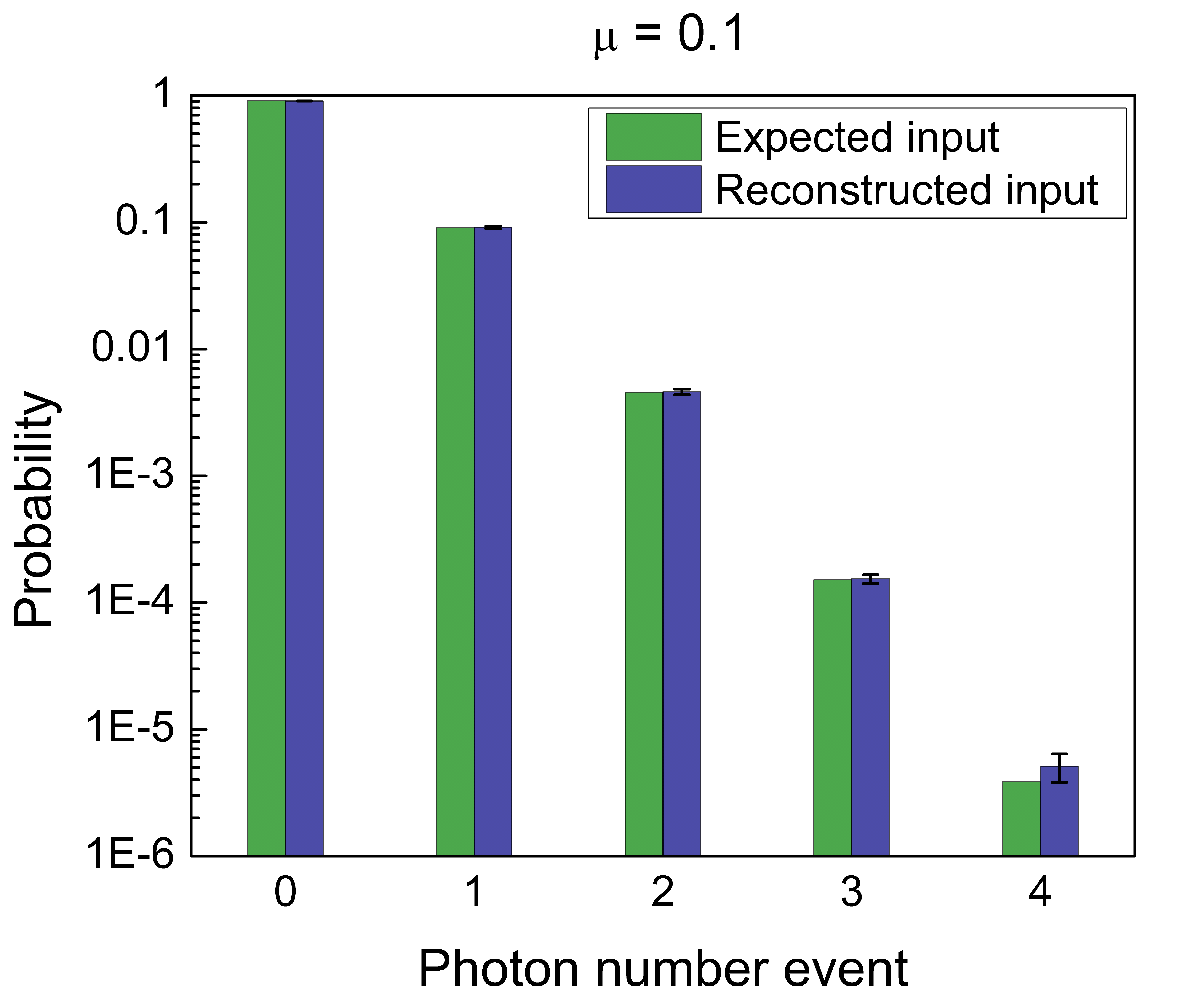}
        \label{fig:mu=0.1}
    \end{minipage}\hfill%
    \begin{minipage}{0.3\textwidth}
        \centering
        \includegraphics[width=\textwidth]{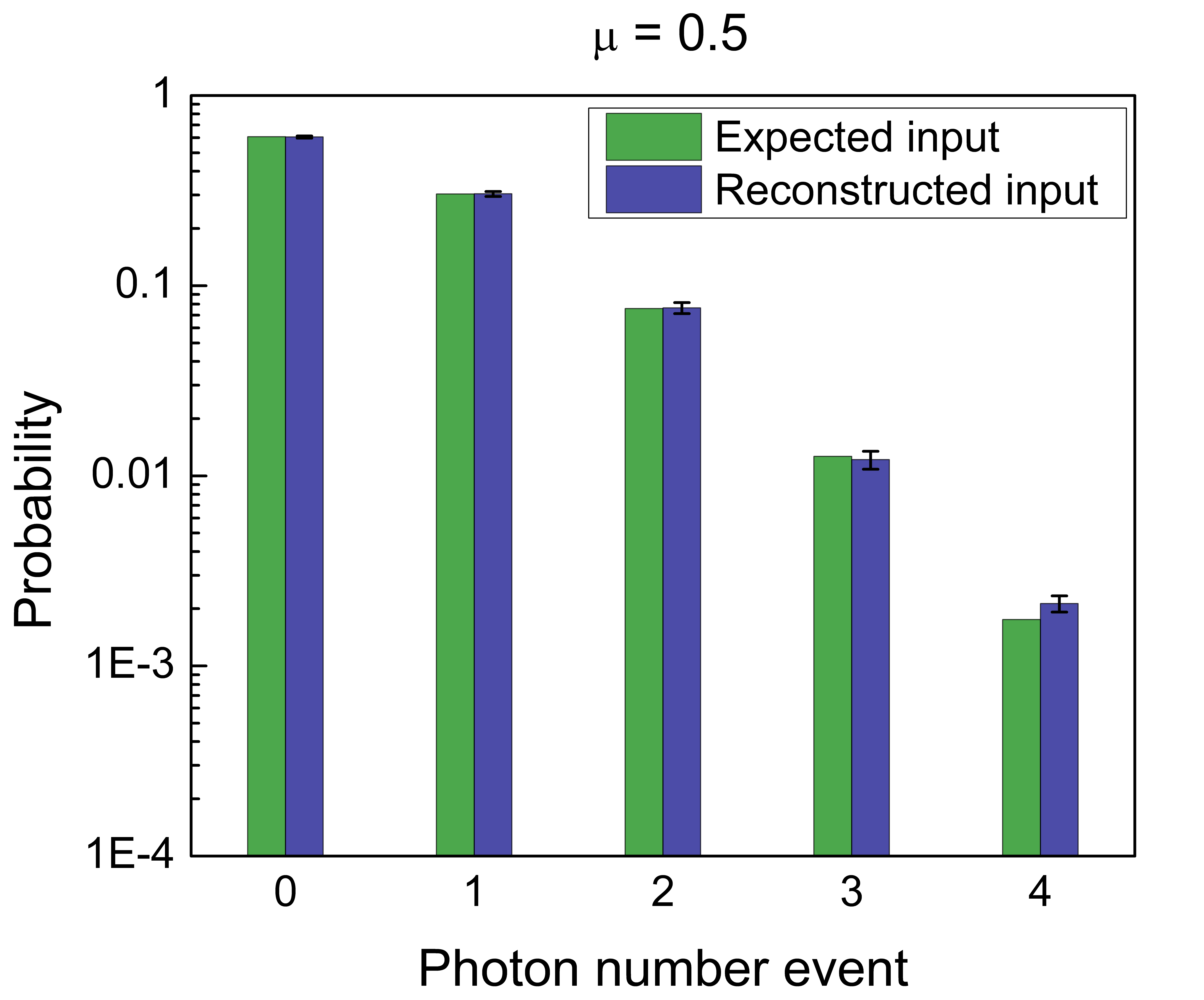}
        \label{fig:mu=0.5}
    \end{minipage}\hfill%
    \begin{minipage}{0.3\textwidth}
        \centering
        \includegraphics[width=\textwidth]{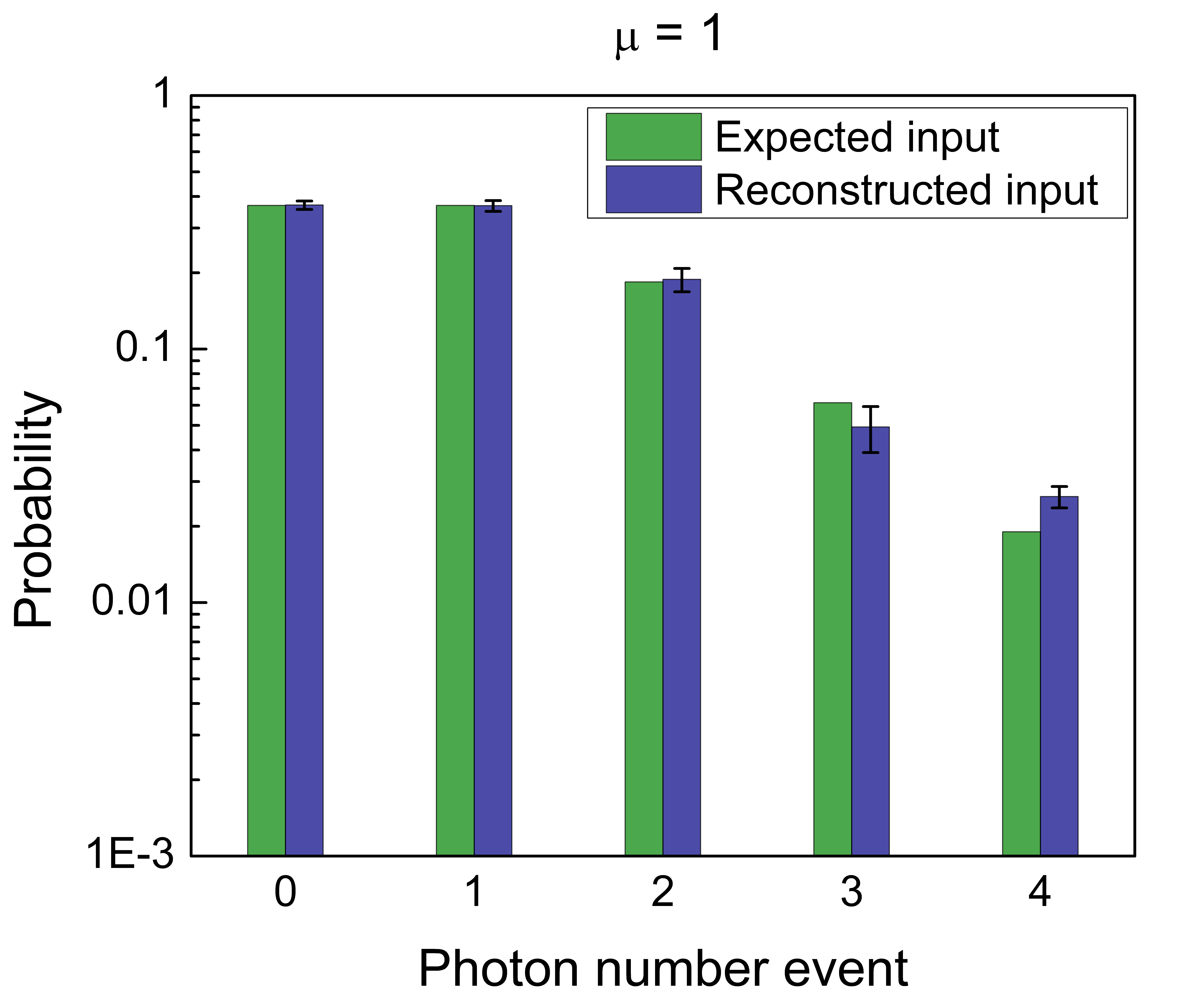}
        \label{mu=1}
    \end{minipage}\hfill
    \caption{Input reconstruction for $\mu=0.1$ (left), $\mu=0.5$ (center), $\mu=1$ (right).}
    \label{fig:input_reconstruction}
\end{figure*}

Such results show how the P-SNSPDs can maintain high SDE, low DCR and low jitter similar to a single meander SNSPDs, displaying a fast RT with more than $50\%$ nominal efficiency after only $10~\nano\second$ and the additional feature of PNR capability.

\subsection{Probabilities matrix}
In the P-SNSPDs architecture, each pixel is connected to the other, thus they cannot be read-out individually. Therefore, it is not possible to compute \textbf{P} directly, since it is a function of the single pixel efficiencies $\{\eta_i\}$.

In order to obtain \textbf{P}, we characterize the PNR capabilities of our detector with a known light source of input statistics \textbf{S} and we record the statistics generated by the detector \textbf{Q}. Then we use an optimization algorithm that finds the pixel efficiencies minimizing the Euclidean norm:
\begin{equation}
    ||\textbf{Q}-{\textbf{P}\textbf{S}}||_2
\end{equation}

As input statistics $S{\left(m\right)}$, we use poissonian light:
\begin{equation}
    S{\left(m\right)} = \frac{\mu ^m}{m!} \mathrm{e}^{-\mu}
\end{equation}
where $\mu$ is the mean photon number per pulse.

The output statistics $Q{\left(n\right)}$ is measured as follows, exploiting the setup in \cref{fig:setup}: 
\begin{enumerate}[(i)]
    \item a pulsed laser (ID3000 from ID Quantique) is triggered at 10~MHz repetition rate, sending a light pulse (22~ps) to the detector and an electrical one to the time tagger (ID900 from ID Quantique). The latter serves as start signal to build histograms for each photon event.
    \item The electrical signal coming from the detector is sent to a 1-to-4 resistive splitter (ZFRSC-4-842-s+ MiniCircuits) and then to the time-tagger as stop signal for the histograms. Different thresholds are set on each input, corresponding to 1-, 2-, 3- and 4-photons event. The counts are taken in a 2~ns window. Counts contributing to a photon-number event are also contained in the counts of a lower photon-number event, hence the real counts for a specific event are: $ c'_n = c_n - c_{n+1}$.
    \item To obtain the 0-photon event, we subtract from the total number of events, the registered ones. The total number of events is computed as ${N_{tot}} = R \cdot t$ where $R$ is the repetition time of the laser and $t$ the time of acquisition which can be set on the ID900.
\end{enumerate}   

Thanks to our design that prevents electrical and thermal crosstalk, and the negligible DCR (the probability to have a dark count in the time window of the experiment is $2.8\times10^{-7}$), we validate the model assumptions of setting to 0 all the $P_{nm}$ elements with $n>m$. 

We acquire several set of data at different $\mu$ using a powermeter and three digital optical attenuators. We span $\mu$ from 0.1 to 2 and the probabilities matrices obtained are in agreement with each others. 

The retrieved pixel efficiencies are $2.48 \pm 0.06 \%$, $35.65 \pm 0.87\%$, $48.62\pm 1.18\%$ and $5.66 \pm 0.14\%$.
The data reflect the Gaussian distribution of light in single-mode fibers, with the outer pixels showing a much lower efficiency compared to the central ones. From those value, we reconstruct \textbf{P} and obtain the fidelity probabilities $P_{nn}$. We limit the matrix dimension to $M=9$ during the optimization process, in order to take into account more than 99.99~\% of the events generated by the poissonian light source. In \cref{eq:POVM1} we report \textbf{P} and the uncertainty values for each $P_{nm}$ element  in \cref{eq:stdPOVM}.

\onecolumngrid
\begin{equation}
    \mathbf{P} = \begin{bmatrix}
        1 & 0.076  & 0.0063 & 0.0005 & 0       & 0      & 0      & 0       & 0      & 0 \\
        0 & 0.924  & 0.5067 & 0.2602 & 0.1354  & 0.0716 & 0.0383 & 0.0207  & 0.0113 & 0.0062 \\
        0 & 0      & 0.487  & 0.6472 & 0.6728  & 0.6482 & 0.6067 & 0.5596  & 0.514  & 0.4712 \\
        0 & 0      & 0      & 0.092  & 0.1858  & 0.2645 & 0.3275 & 0.3777  & 0.4177 & 0.4498 \\
        0 & 0      & 0      & 0      & 0.0058  & 0.0157 & 0.0281 & 0.042   & 0.057  & 0.0767 \\
    \end{bmatrix},
\label{eq:POVM1}
\end{equation}

\begin{equation}
    \sigma_{\mathbf{P}} = \begin{bmatrix}
        0 & 0.0224  & 0.0035    & 0.0004  & 0       & 0      & 0      & 0      & 0      & 0 \\
        0 & 0.0224  & 0.0202    & 0.0178  & 0.0122  & 0.0078 & 0.0049 & 0.003  & 0.0018 & 0.0011 \\
        0 & 0       & 0.0236    & 0.0117  & 0.0026  & 0.003  & 0.0061 & 0.0079 & 0.0088 & 0.0093 \\
        0 & 0       & 0         & 0.0067  & 0.0091  & 0.0095 & 0.0092 & 0.0085 & 0.0077 & 0.007  \\
        0 & 0       & 0         & 0       & 0.0006  & 0.0012 & 0.0018 & 0.0024 & 0.0029 & 0.0035 \\
    \end{bmatrix}.
    \label{eq:stdPOVM}
\end{equation}
\twocolumngrid

The computation of the uncertainties can be found in the \cref{app:appB}.
The $P_{11}$ reflects the SDE obtained with continuous wave laser, proving that the model is consistent with the previous measurement. The drop in fidelity by $P_{33}$ and $P_{44}$ is due to two main factors: 
\begin{itemize}[$\circ$]
    \item the probability that all the photons are split on different pixels decreases, therefore two photons can end up on the same pixel and only one will be detected.
    \item even if all the photons are split on different pixels, all of them have to click to register the corrected event.
\end{itemize}

Once \textbf{P} is computed, we can use it to reconstruct the statistics of unknown light sources. We need to invert \cref{eq:povm} to retrieve \textbf{S}, the light statistics, hence \textbf{P} needs to be square to be invertible. Since the matrix has dimension ${(N+1)}\times M$, we need to reduce its dimension by truncating it. Hence, by neglecting all the $P_{nm}$ elements with $m>N$, we are able to reconstruct only light source statistics in which events with more than $N$-photons are negligible, thus when $\mu$ is low. 
In \cref{fig:input_reconstruction} we report the reconstructed statistics of poissonian light with three different $\mu$ values: for $\mu$ equal to 0.1 and 0.5 there is a very good agreement. At $\mu=1$ the reconstruction starts to deviate from the theoretical input, since events with more than $N$ photons start to be non-negligible. In fact, the detector saturates when more than $N$ photons are impinging on it and we lose information about those higher photon-number events.
To overcome this limitation, P-SNSPDs with a higher number of pixels would allow for a more accurate state reconstruction and provide better fidelity probabilities~\cite{provaznik2020benchmarking}.

\section{Conclusion}
In conclusion, we developed a general model that can be use to characterize the multi-photon absorption probabilities for any multi-pixel SNSPDs. The model is based on the possible combination of clicking pixels for a specific photon-absorption event and we do not take any assumption on the single pixel efficiencies and neither on the light spatial distribution on the detector. We employed the model on a high efficient and fast 4-pixel P-SNSPD that displays properties comparable to commercial single-meander SNSPDs but with the extra-feature of PNR capability and faster recovery time. Thanks to the model, we were able to access the fidelity probabilities of the detector, an additional information that could not be characterized before, and the full $P_{nm}$ matrix that can be used to reconstruct the statistics of unknown light sources.
At the moment we are limited by the low number of pixels, but \psnspd{} seem a promising solution to overcome TES limitations. However, the number of pixels cannot grow indefinitely, since the voltage difference between the $n$-click events would become too small to be resolved. Improvement in the design, such as interleaved nanowires, could improve the fidelity probabilities of \psnspd{}, making them a useful tool in optical quantum computation and quantum metrology. 

\begin{acknowledgments}
We thank Giovanni Resta for useful discussion. L.S. is part of the AppQInfo MSCA ITN which received funding from the EU Horizon 2020 research and innovation program under the Marie Sklodowska-Curie grant agreement No 956071. 
\end{acknowledgments}

\appendix
\section{}
\label{app:appA}
In this appendix, we detail the expression for all the $P_{nm}$ values in the matrix \textbf{P}. Since we are considering that the photons are registered one at the time, it is straightforward to compute the $P_{nm}$ elements keeping the number of clicks fixed while letting the number of incident photons $m$ as a variable. 

\begin{figure}[hbt!]
    \centering
    \includegraphics[width=7cm]{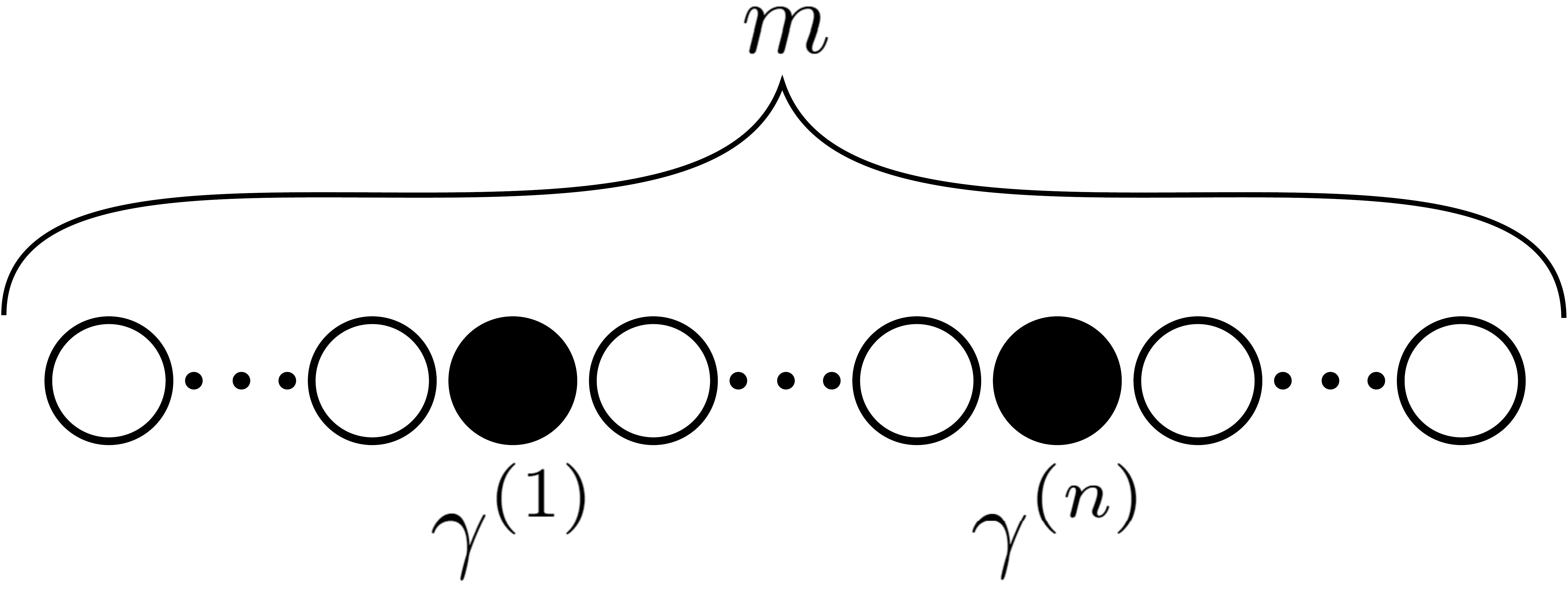}
    \caption{The black circle indicates the position of the photon within the $m$-group that made the detector clicks whereas the white circles represent the non-detected photons.}
    \label{fig:Photon time of arrival_4}
\end{figure}

To generate a 0-click event, the detector must not click $m$ times, therefore the $P_{0m}$ can be written as:
\begin{equation}
    P_{0m} = \biggl( 1 -  \sum_{i=1}^{N} \eta_i \biggl)^{m}.
\end{equation}
The other $P_{nm}$ are constructed as follows (see \cref{fig:Photon time of arrival_4}): 
\begin{itemize}[$\circ$]
    \item the detector miss the first $\gamma^{(1)}-1$ photons with probability $\left( 1 -  \sum_{i=1}^{N} \eta_i \right)^{\gamma^{(1)}-1}$;
    \item it detects the $\gamma^{(1)}$-th photon generating the first click with probability $\sum_{i=1}^{N} \eta_i$;
    \item it can miss photons for other $\gamma^{(2)} -\gamma^{(1)}-1$ times with probability $\left( 1 - \sum_{\substack{j=1\\j \textrm{ active}}}^{N} \eta_j \right)^{\gamma^{(2)}-\gamma^{(1)}-1}$, where we sum over the remaining active pixels $j$ after the first detection.
    \item It detects the $\gamma^{(2)}$-th photon generating the second click with probability $\sum_{\substack{j=1\\j \textrm{ active}}}^{N} \eta_j$ where we sum over the remaining active pixels $j$ after the first detection;
    \item we iterate such operation depending on the $n$-click event we consider;
    \item after the last photon is detected, the remaining photons will be missed with probability $\left( 1 - \sum_{\substack{k=1\\k \textrm{ active}}}^{N} \eta_k \right)^{m-\gamma^{(n)}}$
\end{itemize}
By enumerating all the possible configurations of absorbed photons within the $m$-group and all the possible combinations of clicking pixels, we can obtain the $P_{nm}$ elements.

\onecolumngrid
\begin{equation}
    P_{1m} =\sum_{1\leq\gamma^{(1)}\leq m}^{m} \Bigg\{ \biggl( 1 -  \sum_{i=1}^{N} \eta_i \biggl)^{\gamma^{(1)}-1} \cdot \sum_{i=1}^{N} \eta_i \cdot \biggl( 1 - \sum_{\substack{j=1, \\ j\textrm{ active}}}^{N} \eta_j \biggl)^{m-\gamma^{(1)}} \Bigg\},
\end{equation}
\begin{equation}
\begin{aligned}
    P_{2m} {}= 
    & \sum_{1\leq\gamma^{(1)}<\gamma^{(2)}\leq m}^{m} \Bigg\{ \biggl( 1 -  \sum_{i=1}^{N} \eta_i \biggl)^{\gamma^{(1)}-1} \cdot \biggl\{ \sum_{i=1}^{N} \eta_i \cdot \biggl( 1 - \sum_{\substack{j=1, \\ j\textrm{ active}}}^{N} \eta_j  \biggl)^{\gamma^{(2)}-\gamma^{(1)}-1} \cdot \\
    & \biggl[ \sum_{\substack{j=1, \\ j\textrm{ active}}}^{N} \eta_j \cdot \biggl( 1 - \sum_{\substack{k=1, \\ k\textrm{ active}}}^{N} \eta_k \biggl)^{m-\gamma^{(2)}} \biggl] \biggl\} \Bigg\},
    \end{aligned}
\end{equation}

\begin{equation}
\begin{aligned}
    P_{3m} {}= 
    &\sum_{1\leq\gamma^{(1)}<\gamma^{(2)}<\gamma^{(3)}\leq m}^{m} \Bigg\{ \biggl( 1 - \sum_{i=1}^{N} \eta_i \biggl)^{\gamma^{(1)}-1} \cdot \biggl\{ \sum_{i=1}^{N} \eta_i \cdot \biggl( 1 - \sum_{\substack{j=1, \\ j\textrm{ active}}}^{N} \eta_j \biggl)^{\gamma^{(2)}-\gamma^{(1)}-1} \cdot \\ 
    & \biggl\{ \sum_{\substack{j=1, \\ j\textrm{ active}}}^{N} \eta_j \cdot \biggl( 1 - \sum_{\substack{k=1, \\ k\textrm{ active}}}^{N} \eta_k \biggl)^{\gamma^{(3)}-\gamma^{(2)}-1} \cdot \biggl[ \sum_{\substack{k=1, \\ k\textrm{ active}}}^{N} \eta_k \cdot \biggl( 1 - \sum_{\substack{h=1, \\ h\textrm{ active}}}^{N} \eta_h \biggl)^{m - \gamma^{(3)}} \biggl] \biggl\} \biggl\} \Bigg\},
\end{aligned}
\end{equation}

\begin{equation}
\begin{aligned}
    P_{4m} {}= 
    & \sum_{1\leq\gamma^{(1)}<\gamma^{(2)}<\gamma^{(3)}<\gamma^{(4)}\leq m}^{m} \Bigg\{ \biggl( 1 - \sum_{i=1}^{N} \eta_i \biggl)^{\gamma^{(1)}-1}
    \cdot \biggl\{ \sum_{i=1}^{N} \eta_i \cdot \biggl( 1 - \sum_{\substack{j=1, \\ j\textrm{ active}}}^{N} \eta_j \biggl)^{\gamma^{(2)}-\gamma^{(1)}-1} \cdot \\
    & \biggl\{ \sum_{\substack{j=1, \\ j\textrm{ active}}}^{N} \eta_j \cdot \biggl( 1 - \sum_{\substack{k=1, \\ k\textrm{ active}}}^{N} \eta_k \biggl)^{\gamma^{(3)}-\gamma^{(2)}-1} 
    \cdot \biggl[ \sum_{\substack{k=1, \\ k\textrm{ active}}}^{N} \eta_k \cdot \biggl( 1 - \sum_{\substack{h=1, \\ h\textrm{ active}}}^{N} \eta_h \biggl)^{\gamma^{(4)} - \gamma^{(3)}-1} \cdot \\
    & \biggl[\sum_{\substack{h=1, \\ h\textrm{ active}}}^{N} \eta_h \cdot \biggl( 1 - \sum_{\substack{l=1, \\ l\textrm{ active}}}^{N} \eta_l \biggl)^{m- \gamma^{(4)}} \bigg] \biggl] \biggl\} \biggl\} \Bigg\}.
\end{aligned}
\end{equation}
\twocolumngrid

\section{}
\label{app:appB}
The sources of error in our set-up come from the power meter (PM), the 99/1 optical coupler (OP) and the three variable attenuators (AT). 
To assess the uncertainty on the power meter, we measure the laser power with 5 different power meters of the same model and we find that the standard deviation is 2.52\%.

For the optical coupler, we measure the ratio between the two output power several times and the uncertainty is 0.19\%. 

The attenuators are recalibrated each measurement using the powermeter by measuring the ratio between $\frac{P_{AT}}{P_0}$, where $P_{AT}$ is the power value when the attenuation is on, and $P_0$ is the power value when the attenuation is zero. Therefore, the only contribution for the variable attenuators is represented by their repeatability and is measured to be 0.12\%.

To compute the overall uncertainty on the total number of photons per second $N_\gamma$ sent on the detector, we use the equation presented in Ref. \cite{caloz2018high}:

\begin{equation}
    \left( \frac{\sigma_{N_\gamma}}{N_\gamma}\right)^2 = \left( \frac{\sigma_{PM}}{P_{PM}}\right)^2 + \left( \frac{\sigma_{OP}}{R_{OP}} \right)^2 + 3 \cdot\left( \frac{\sigma_{AT}}{AT}\right)^2
\end{equation}
\noindent where $P_{PM}$ is the power read by the power meter, $R_{OP}$ is the measured value of the 99/1 optical coupler, AT is the attenuation value and the $\sigma_i$ the associated uncertainties.

Since we couldn't use error propagation throughout the optimization algorithm we carried out Monte Carlo simulations to estimate the error on each $P_{nm}$ elements.  Therefore, we construct several simulated set of data, with the following method:
\begin{enumerate}[(i)]

    \item to obtain the new number of counts for a specific $n$-click event, we randomly toss a coin $N_t$ times with probability of success given by the experimental photon-number distribution $Q{(n)}$ recorded by the detector. The number of successful events correspond to the new number of $n$-click events. The 0-click event are obtained by subtracting from $N_t$ the 1-, 2-, 3- and 4-click events, to ensure that $Q'{(n)}$ is normalized.
    \item The error associated to the input light statistics $S{(m)}$ is given by the $\sigma_{N_\gamma}$ and it was found to be 2.53\%. Thus, to construct the new input light statistics $S'{(m)}$, we randomly choose a mean photon number per pulse $\mu$ from a Gaussian distribution centered in $\mu$ with standard deviation $\sigma =0.0253 \cdot \mu$. This new $\mu$ is then use to construct the poissonian light input statistic $S'{(m)}$.
\end{enumerate}

$Q'{(n)}$ and $S'{(m)}$ are feed to the optimization algorithm that output the matrix \textbf{P}. We compute an average between the retrieved \textbf{P} and we use the standard deviation as the uncertainty on each $P_{nm}$ value.

\end{document}